\documentclass[twoside]{article}
\usepackage{graphicx,rotating}
\usepackage{fancyhdr}
\usepackage{multicol}
\usepackage{styl-ap}

\begin{document}
\title{Inner Disk Structure of Dwarf Novae in the Light of X-ray Observations}
\author{\c{S}\"{o}len Balman\work{1}}
\workplace{
Middle East Technical University, Department of Physics,
Inonu Bulvar{\i}, Ankara, 06531, Turkey}
\mainauthor{solen@astra.physics.metu.edu.tr}
\maketitle

\begin{abstract}%
   Diversity of the X-ray observations of dwarf nova are still not
   fully understood. I review the X-ray spectral characteristics
   of dwarf novae during the quiescence in general explained by
   cooling flow models and the outburst spectra that show hard X-ray
   emission dominantly with few sources that reveal soft X-ray/EUV blackbody emission. 
   The nature of aperiodic
   time variability of brightness of dwarf novae shows band limited noise, which
   can be adequately described in the framework of the model of propagating
   fluctuations. The frequency of the break (1-6 mHz) indicates  inner disk
   truncation of the optically thick disk with a range of radii (3.0-10.0)$\times$10$^{9}$ cm. 
   The $RXTE$ and optical (RTT150) data of SS Cyg in outburst and quiescence
   reveal that the inner disk radius moves towards the white  dwarf and
   receeds as the outburst declines to quiescence. A preliminary
analysis of SU UMa indicates a similar behaviour. In addition, 
I find that the outburst spectra of WZ Sge shows two component spectrum 
of only hard X-ray emission, one of which may be fitted with a power law 
suggesting thermal
Comptonization occuring in the system. Cross-correlations
   between the simultaneous UV and X-ray light curves ($XMM-Newton$) of five DNe in 
quiescence  show time lags in
   the X-rays of 96-181 sec consistent with travel time of matter from a
   truncated inner disk to the white dwarf surface.  
   All this suggests that dwarf novae and other plausible nonmagnetic 
   systems have truncated
   accretion disks indicating that the disks may be partially evaporated
   and the accretion may occur through hot (coronal) flows in the disk.

\end{abstract}


\begin{multicols}{2}
\section{Introduction}

Dwarf novae (DNe) are a class of cataclysmic variables
(CVs) which are interacting compact binaries
in which a white dwarf (WD, the primary star) accretes matter
and angular momentum from a main (or post-main) sequence star (the secondary)
filling its Roche-lobe. The matter is transferred
by means of an {\it accretion} disk that is assumed to reach all the way to the WD surface. 
Ongoing accretion at
a low rate (quiescence) is interrupted every few weeks to months or sometimes with longer durations by
intense accretion (outburst) of days to weeks where $\dot{\rm M}$ increases 
(see Warner 1995 for a review).

The material in the inner disk of nonmagnetic CVs initially moving with Keplerian velocity dissipates its
kinetic energy in order to accrete onto the slowly rotating WD creating a boundary layer (BL)
(see Mauche 1997, Kuulkers et al. 2006 for an overview).
Standard accretion disk theory predicts half of the accretion luminosity to originate
from the disk in the optical and ultraviolet (UV) wavelengths and the other half to emerge from the
boundary layer as X-ray and/or extreme UV (EUV)/soft X-ray emission which may be summerized as
L$_{BL}$$\sim$L$_{disk}$=GM$_{WD}$$\dot M_{acc}$/2R$_{WD}$=L$_{acc}$/2 (Lynden-Bell $\&$ Pringle 1974,
Godon et al. 1995). During low-mass accretion rates, $\dot M_{acc}$$<$10$^{-(9-9.5)}$M$_{\odot}$,
the boundary layer is optically
thin (Narayan $\&$ Popham 1993, Popham 1999) emitting mostly in the hard X-rays (kT$\sim$10$^{(7.5-8.5)}$ K).
For higher accretion rates , $\dot M_{acc}$$\ge$10$^{-(9-9.5)}$M$_{\odot}$, the boundary layer is expected to be
opticallly thick (Popham $\&$ Narayan 1995) emitting in the soft X-rays or EUV (kT$\sim$10$^{(5-5.6)}$ K).
The transition between an optically thin and an optically thick
boundary layer, also depends on the mass of the white dwarf (also rotation) and on the alpha
viscosity parameter. 

\section{X-ray Observations of Dwarf Novae in Quiescence and Outburst}

The quiescent X-ray spectra are mainly characterised with a multi-temperature isobaric cooling flow
model of plasma emission at T$_{max}$=6-55 keV with accretion rates of 
10$^{-12}$-10$^{-10}$ M$_{\odot}$ yr$^{-1}$.
The X-ray line spectroscopy indicates narrow emission lines (brightest OVIII K$\alpha$) 
and near solar abundances, with a 6.4 keV line expected to be due to reflection from the surface of the WD. 
The detected Doppler broadening in lines during quiescence is $<$750 km s$^{-1}$ at sub-Keplerian velocities in
the boundary layer with 
electron densities $>$10$^{12}$ cm$^{-3}$ 
(see  Baskill et al. 2005, Kuulkers et al. 2006, Rana et al. 2006, Pandel et al. 2005, 
Balman et al. 2011, Balman 2012). The total X-ray luminosity during quiescence is 10$^{29}$ -10$^{32}$ erg s$^{-1}$.
Lack of BL emission in the X-rays have been suggested due to low L$_x$/L$_{disk}$ ratio (see Kuulkers et al. 2006). 
It has been suggested for DN in quiescence that if the WD emission is removed and 
some disk truncation is allowed, this ratio is $\sim$1 (Pandel et al. 2005).

DN outbursts are brightenings of the accretion disks as a result of thermal-viscous instabilities summerized
in the DIM model (Disc Instability Model; Lasota 2001,2004).
During the outburst stage, DN X-ray spectra differ from the quiescence since the
accretion rates are higher (about two orders of magnitude), the BL is expected to be optically thick emitting
EUV/soft X-rays (Lasota 2001, see the X-ray review in Kuulkers et al. 2006). On the other hand,
soft X-ray/EUV temperatures in a range 5-25 eV are detected from only about five systems 
(e.g., Mauche et al. 1995, 
Mauche $\&$ Raymond 2000, Long et al. 1996, Mauche 2004, Byckling et al. 2009). As a second and more
dominantly detected emission
component, DN show hard X-ray emission during the outburst stage however, at a lower flux level and X-ray
temperature compared with the quiescence (e.g., WW Cet $\&$ SU UMa: Collins \& Wheatley 2010, 
Fertig et al. 2011, SS Cyg: Wheatley et al. 2003, MacGowen et al. 2004, Ishida et al. 2009).
On the other hand, some DN show increased level of X-ray emission  
(GW Lib \& U Gem: Byckling et al. 2009, Guver et al. 2006). 
The total X-ray luminosity during outburst is 10$^{30}$ -10$^{34}$ erg s$^{-1}$. 
The grating spectroscopy of the outburst 
data indicate  large widths for lines with velocities in excess of 1000 km s$^{-1}$ 
mostly of H and He-like emission lines (C,N,O,Ne, Mg, Si, Fe, ect.) 
(Mauche 2004, Rana et al. 2006, Okada et al. 2008). A characteristic of some DN outburst light curves are the
UV and X-ray delays in rise to outburst (w.r.t. optical) indicating possible disk truncation
((Meyer $\&$ Meyer-Hofmeister 1994 and references therein) .
These delays are a matter of several hours (upto a day) that need dedicated simultaneous 
multi-wavelength observations.
During outburst no  eclipses are detected in the eclipsing systems (particularly of soft X-ray emission)  
or no distinct orbital variations are seen. (e.g., Pratt et al. 1999, Byckling et al. 2009).
Note that both in SS Cyg (Mc Gowan et al. 2004) and in SU UMa (Collins \& Wheatley 2010) 
the X-ray flux in between outbursts have been found to decrease as opposed to expectations of the DIM model.
 
WZ Sge is a short period SU UMa type dwarf nova with long interoutburst interval of 20-30 yrs.
The system is believed to show no normal outbursts but only superoutbursts.
The most recent outburst in July-August 2001 (previous outbursts in 1978, 1946 and 1913)
was observed from ground and space over several different wavelength bands including
the X-ray wavelengths and the EUV using the 
$Chandra$ Observatory (see Wheatley \& Mauche 2005, Kuulkers et al. 2002).  
There has been three LETG (Low Energy Transmission Grating) and three ACIS-S 
(Advanced 
Camera Imaging Specreometer) observations obtained in the 
continuous-clocking (CC) mode.  I report a preliminary analysis on the spectral fitting of the 
ACIS-S CC-mode data. 

\begin{myfigure}
\begin{turn}{-90}
\resizebox{60mm}{!}{\includegraphics{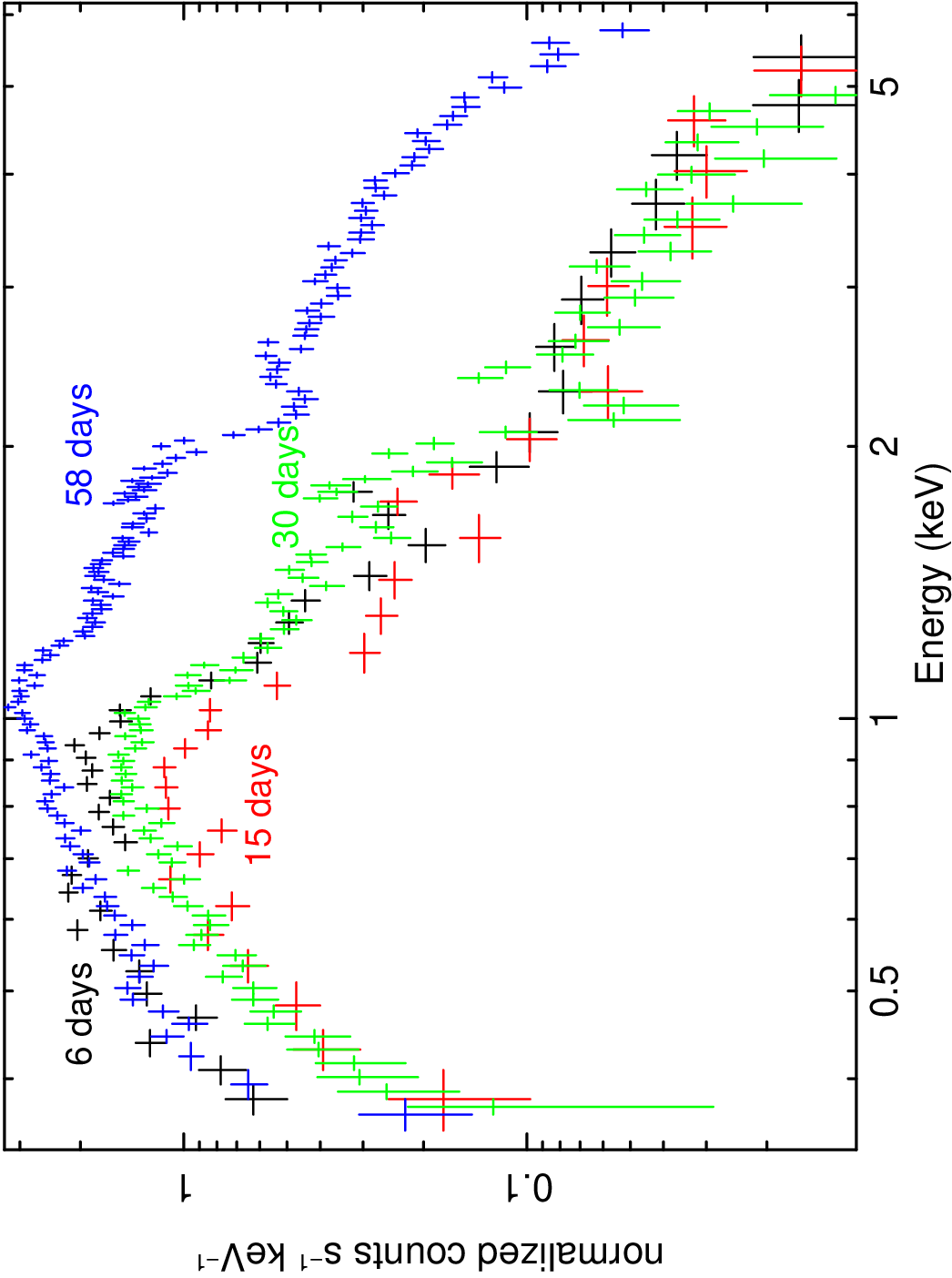}}
\end{turn}
\caption{Chandra X-ray outburst spectra of WZ Sge in 0.3-10.0 keV range obtained
with ACIS-S (CC-mode). The observation days after the
outburst are labeled on each spectrum.}
\end{myfigure}

The spectra and response files
were prepared in a standard way (task {\sc specextract} is used) by choosing the source and the
background photons using a suitable extraction box (10 arcsec in size) on the data strip using 
CIAO 4.4 and the CALDB 4.4.8 (see http://cxc.harvard.edu/ciao/). 
For further analysis, HEASOFT version 6.13 is utilized. 
The spectra were fitted with the multi-temperature plasma emission 
models (e.g., CEVMKL in XSPEC, see http://heasarc.nasa.gov/xanadu/xspec/) 
according to the expections from earlier work
of DN analysis in quiescence and outburst. 
CEVMKL model is a multi-temperature plasma emission model (built from mekal code, Mewe et al. 1986) where 
Emission measures follow a power-law in temperature (dEM = (T/Tmax)$^{\alpha-1}$ dT/Tmax).
The reduced $\chi^2$ of the
fits were above a value of 2 with either a single MEKAL model or CEVMEKL alone
(the $\alpha$ parameter of the power law 
distribution of temperatures is set free) for days 6, 15, 30 after outburst.
I added a second component of a power law which reduced the $\chi^2$ values
to desirable levels. In general, the plasma has variable abundance of N,O,Ne,Fe, and S.
The spectral parameters for day 6 after the outburst were
kT$_{max}$=0.7-1.3 keV with a photon index of $\Gamma$=0.8-1.4. 
The X-ray flux of the thermal CEVMKL component was 1.1$\times$10$^{-11}$ erg
s$^{-1}$ cm$^{-2}$ and L$_{cevmkl}$ is 2.7$\times$10$^{30}$ erg s$^{-1}$ 
where the power law component has  5.8$\times$10$^{-12}$ erg s$^{-1}$ 
cm$^{-2}$ and L$_{power}$ is 1.3$\times$10$^{30}$ erg s$^{-1}$
(43.5 pc is assumed, Harrison et al. 2004).
The parameters
for day 15 were kT$_{max}$=0.5-1.5 keV and $\Gamma$=0.2-1.1. Between these two
dates the X-ray flux of the CEVMKL component decreased by 2 and
the power law component inceased by about 1.4\ . For day 30
kT$_{max}$=1.2-3.0 keV and  $\Gamma$=1.5-2.0. Therefore, the X-ray emitting
region gets hotter and the photon index decrease. The X-ray fluxes recover 
to day 6. The final observation on day 58 which is almost quiescence after the
outburst shows kT$_{max}$=26-46 keV and no power law emission is
detected. 
The X-ray flux is the highest 4.4$\times$10$^{-11}$ erg 
s$^{-1}$ cm$^{-2}$ and L$_{x}$ is 1.1$\times$10$^{31}$ erg s$^{-1}$ 
(at 43.5 pc). 
All ranges correspond to 90\% confidence level errors.
The neutral hydrogen absorption has stayed at the interstellar
level in all fits within errors. No optically thick blackbody 
emission is detected and only X-ray suppression of the quiescent emission is
observed just like some other DN. The X-ray temperatures on days 6-30 are in 
good agreement with the LETG results as well. 
A model of three additive MEKALs 
yield acceptable fits on days 6-30, however allowing for very high X-ray tempartures.
A partial covering absorber model improves the CEVMKL fits, however the high intrinsic absorption
when modeled yields inconsistent count rates for the LETG observations.    
 
\section{Inner Disk structure using Eclipse Mapping Techniques}

Flickering is a fast intrinsic brightness scintilation occurring on time scales from 
seconds to minutes (e.g., amplitudes of 0.01 - 1 mag in the optical). It is observed in 
all accreting sources. Eclipse mapping methods have been used to reproduce the spatial 
distribution of flickering in CVs in the optical and UV wavelengths. 
Some eclipse mapping studies of
flickering in quiescent dwarf novae indicate that mass accretion rate
diminishes by a factor of 10-100 and sometimes by 1000 in the inner regions
of the accretion disks as revealed by the brightness temperature calculations 
which do not find the expected R$^{-3/4}$ radial dependence of brightness temperature
in standard steady-state 
constant mass accretion rate disks (see e.g., Z Cha and OY Car: Wood 1990, 
V2051 OPh: Baptista \& Bortoletto 2004, V4140 Sgr: Borges \& Baptista 2005). 
Biro (2000) finds that this flattening in the brightness
temperature profiles may be lifted by introducing disk truncation in the quiescent state 
(e.g., r $\sim$ 0.15R$_{L1}$ $\sim$4$\times$10$^{9}$ cm; DW UMa, a nova-like). 
A comprehensive UV modeling of accretion disks
at high accretion rates in 33 CVs including many nova-likes and old novae
(Puebla et al. 2007) indicate an extra component from an extended optically
thin region (e.g., wind, corona/chromosphere) evident from
the strong emission lines and the P Cygni profiles. This study also indicates that the mass accretion 
rate may be decreasing 1-3 orders of magnitude in the inner disk region
(divergence starts r$\le$ afew $\times$10$^{10}$ cm towards the WD). 

\section{Inner Disk structure of Dwarf Novae}

The truncation of the optically thick accretion disk in DNe in
quiescence was invoked as a possible explanation for the time lags between the optical and UV 
fluxes in the
rise phase  of the outbursts (Meyer $\&$ Meyer-Hofmeister 1994), 
and for some implications of the DIM (see Lasota 2001, 2004) 
or due to the unusual shape of
the optical spectra or light curves of DNe (Linell et al. 2005, Kuulkers et al. 2011).
Theoretical support for such a two-phase
flow was given by a model of the disk evaporation of (Meyer $\&$ Meyer-Hofmeister 1994).
This model was later elaborated to show that the disk evaporation (coronal ``syphon" flow) may 
create optically thick-optically thin transition regions at various distances from the WD 
(Liu et al. 1997, Mineshige et al. 1998).

\subsection{Propagating Fluctuations Model}

Another diagnostic tool proposed to study the inner disk structure
in accreting objects is the aperiodic variability of brightness of sources in the X-rays.
While the long time-scale variability might be created in the outer parts of the accretion disk
(Warner $\&$ Nather 1971), the relatively fast time variability
(at $f>$few mHz) originates in the inner parts of
the accretion flow (Bruch 2000; Baptista $\&$ Borteletto 2004).
Properties of this noise is similar to that of the X-ray binaries with neutron stars and black holes.
Now, the widely accepted
model of origin for this aperiodic flicker noise is a model of propagating fluctuations
(Lyubarskii 1997, Churazov et al. 2001, Uttley $\&$ McHardy 2001, Revnivtsev et al. 2009,2010,
Uttlet et al. 2011, Scaringi et al. 2012).The modulations of the light are created by variations in
the instantaneous value of the mass accretion rate in the region of the energy release. 
These variations in the mass accretion rate, in turn, are inserted into the flow at all Keplerian radii of 
the accretion disk due to the stochastic nature of
its viscosity and then transferred toward the compact object. Thus, variations are on dynamical timescales.
This model predicts that the truncated accretion disk should lack some part of its variability at
high Fourier frequencies, i.e. on the time scales shorter than a typical time scale of variability.

\subsection{Broad-band Noise of Dwarf Novae}

A recent work by Balman \& Revnivtsev (2012) have used the broad-band noise characteristic
of selected DN in quiescence (only one in outburst: SS Cyg)
and studied the inner disk structure and disk truncation via propagating fluctuations model.
The power spectral densities (PDS) expressed were calculated in terms of the fractional 
rms amplitude squared
following from (Miyamoto et al. 1991). The light curves were divided into segments using 1-5 sec binning
in time and several
PDS were averaged to create a final PDS of the sources.
The white noise levels were subtracted hence
leaving us with the rms fractional variability of the time series in units of $(\rm{rms}/\rm{mean})^2$/Hz.
Next, the rms fractional variability per hertz was multiplied with the frequencies to yield
${\nu} P_{\nu}$ versus ${\nu}$.
For the model fitting a simple analytical model is used 

$$ P({\nu}) \propto {\nu}^{-1} \left(1 + \left(\frac{\nu}{\nu_0}\right)^4 \right)^{-1/4} $$,

\noindent
which was proposed to describe the power spectra of sources with truncated
accretion disks (see Revnivtsev et al. 2010,2011).
The broad-band noise structure of the Keplerian disks often
show $\propto$ $f^{-1\dots-1.3}$ dependence on frequency
(Churazov et al. 2001, Gilfanov et al. 2005), and this
noise will show a break if the optically thick disk truncates as the Keplerian motion
subsides.

Balman \& Revnivtsev (2012) show that for five
DN systems, SS Cyg, VW Hyi, RU Peg, WW Cet and T leo, the UV and X-ray power
spectra show breaks in the variability with break frequencies in a range 1-6 mHz,
indicating inner disk truncation in these
systems. The truncation radii for DN are calculated in a range $\sim$(3-10)$\times$10$^{9}$ cm
including errors (see Table 2 in Balman \& Revnivtsev 2012).

The same authors used the archival $RXTE$ data of SS Cyg in quiescence and
outburst listed in Table 1 of their paper to derive the
broad-band noise of the source in different states (i.e., accretion rates). For the outburst phases,
authors investigated times during the X-ray suppression
(e.g., the X-ray dips; optical peak phases of the outburst) and the X-ray peak.
This, in turn is expected to indicate the motion of the
flow in the inner regions of the disk and the geometry of the inner disk.
Authors show that the disk moves towards the white dwarf during the optical peak to
$\sim$ 1$\times$10$^{9}$ cm and receeds
as the outburst declines to quiescence to 5-6$\times$10$^{9}$ cm. This is shown
for a CV, observationally, for the
first time in the X-rays (see Figure 2). This is also supported with the optical data analysis
of SS Cyg in quiescence and outburst (Revnivtsev et al. 2012) 

\begin{myfigure}
\begin{turn}{-90}
\resizebox{53mm}{!}{\includegraphics{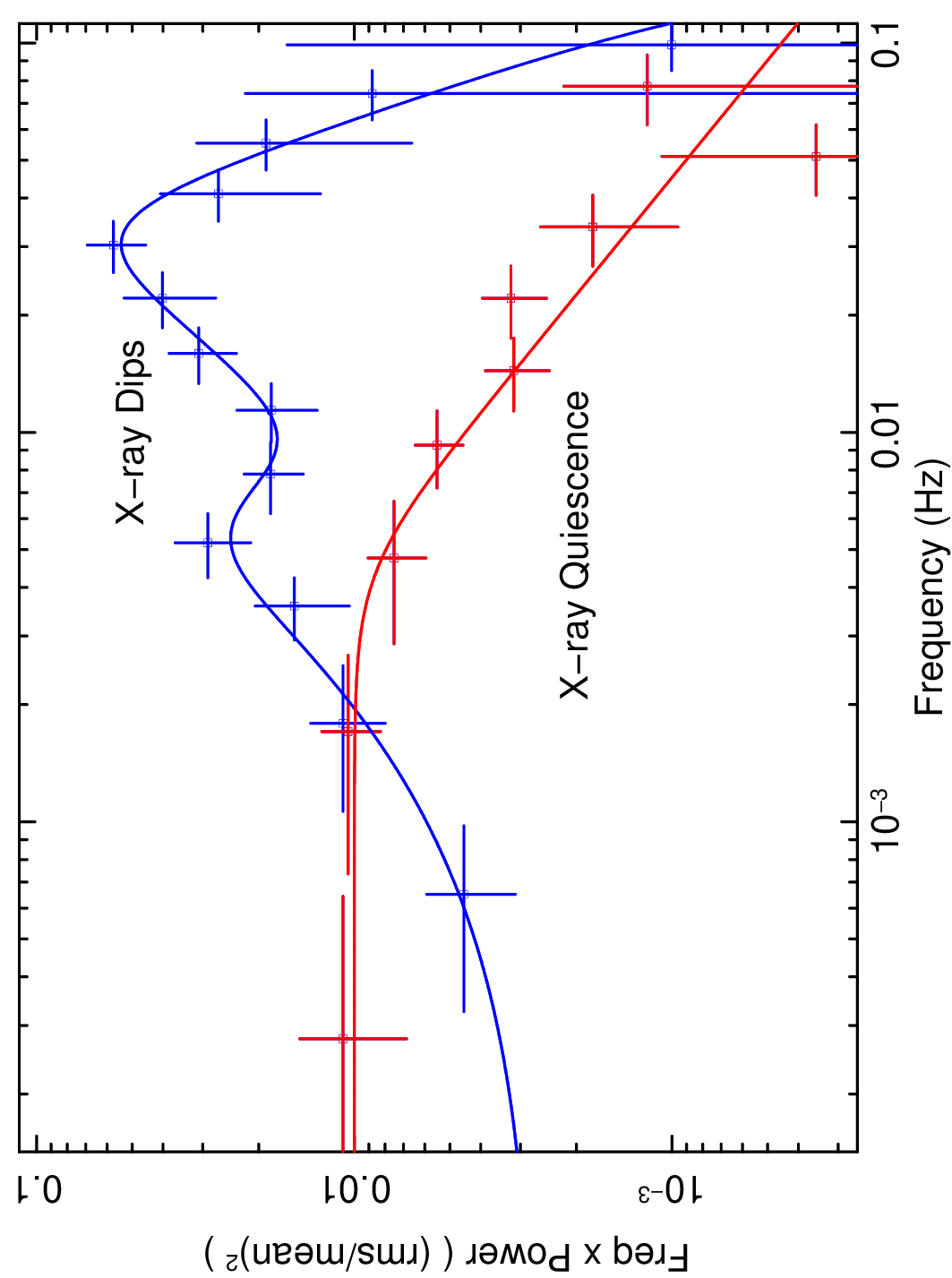}}
\end{turn}
\caption{The $RXTE$ PDS of SS Cyg in quiescence and during the optical peak of outburst (top).
The solid lines show the fit with the propagating fluctuations model along with
two Lorentzians.}
\end{myfigure}

\subsection{X-ray and UV Light Curve Cross-correlations}

Balman \& Revnivtsev (2012)  calculated the cross-correlation between the two
simultaneous light curves (X-ray and UV), using the archival $XMM$-$Newton$ 
EPIC pn and OM data utilizing {\sc HEAsoft} task {\sc crosscor}.
To obtain the CCFs (cross-correlation functions), datasets were divided 
into several pieces using 1-5 sec binning in the 
light curves and fitted the resulting CCFs with double Lorentzians since it was 
necessary for adequate fitting (see the paper for details).
The CCFs for all the DNe show  clear asymmetry indicating that some part of
the UV flux is leading the X-ray flux. In addition, they detect a strong peak
near zero time lag for RU Peg, WW Cet and T Leo, suggesting a significant zero-lag 
correlation between the X-rays and the UV light curves.

The same authors also calculated the cross-correlation  between
the simultaneous UV and X-ray light curves by subtracting the zero time
lag components in the five DNe PDS, yielding more correct time lags consistent with
delays in the X-rays of 96-181 sec (see the paper on details of the modeling). 
The lags occur such that the UV variations lead X-ray variations
which shows that  as the accreting material travels onto the WD, the variations are
carried from the UV into the X-ray emitting region. 
Therefore, the long time lags of the order of minutes 
can be explained by the travel time of matter (viscous flow) from a truncated
inner disk to the white dwarf surface.
Zero time lags ($\sim$ light travel time) indicate irradiation effects in these systems
since we do not have resolution better than 1 sec. 

\begin{myfigure}
\begin{turn}{-90}
\resizebox{43mm}{!}{\includegraphics{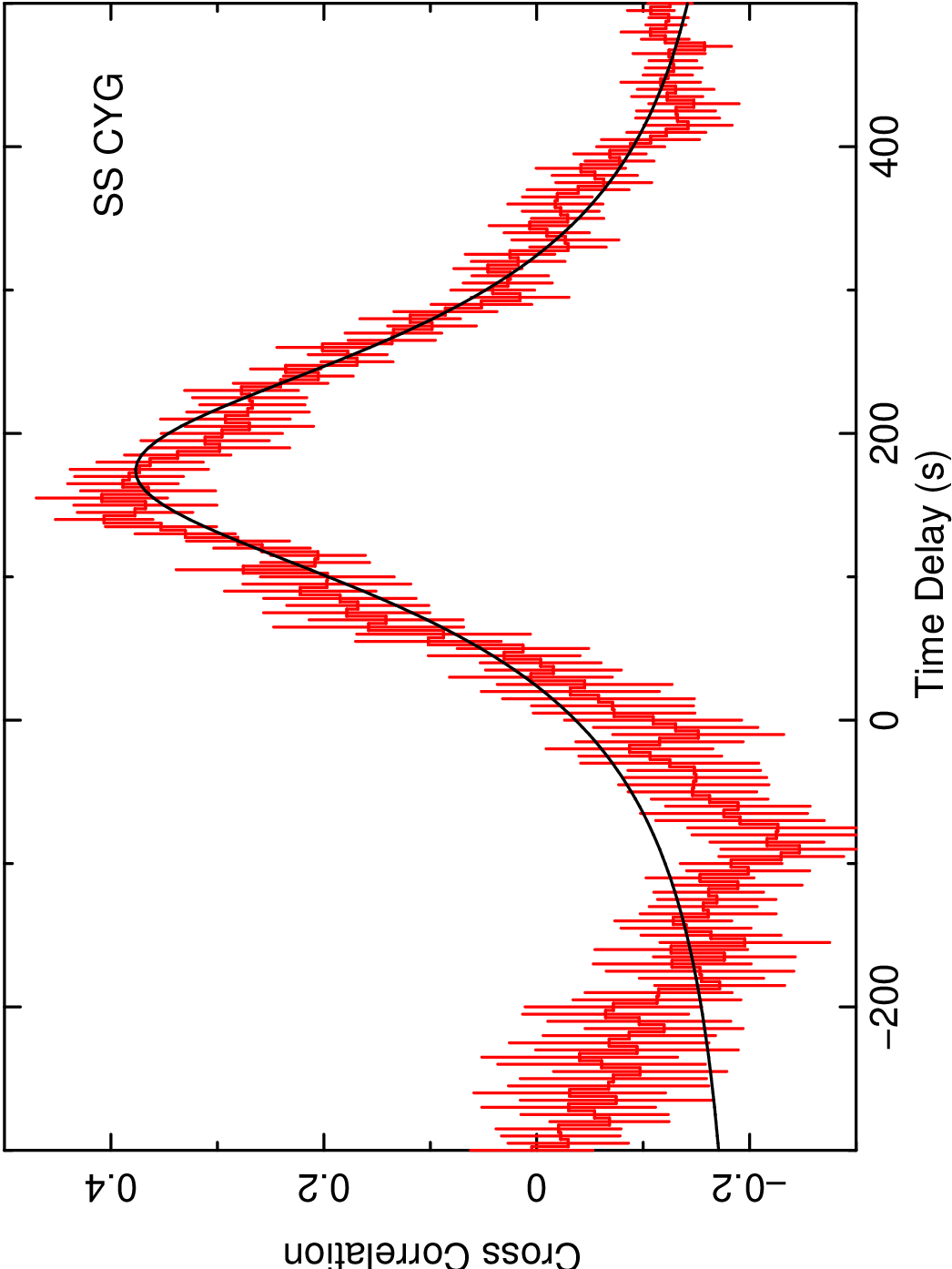}}
\resizebox{40mm}{!}{\includegraphics{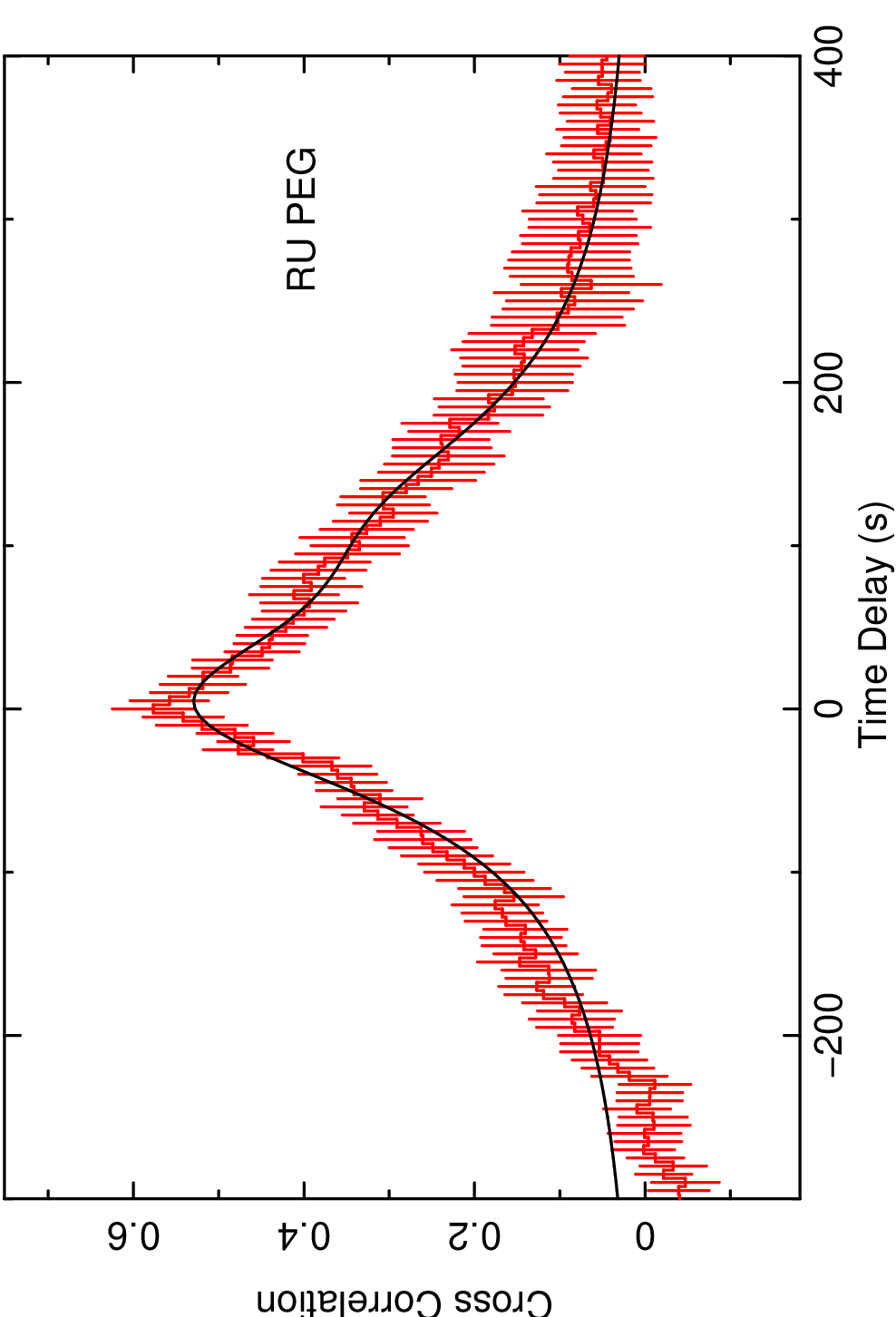}}
\resizebox{40mm}{!}{\includegraphics{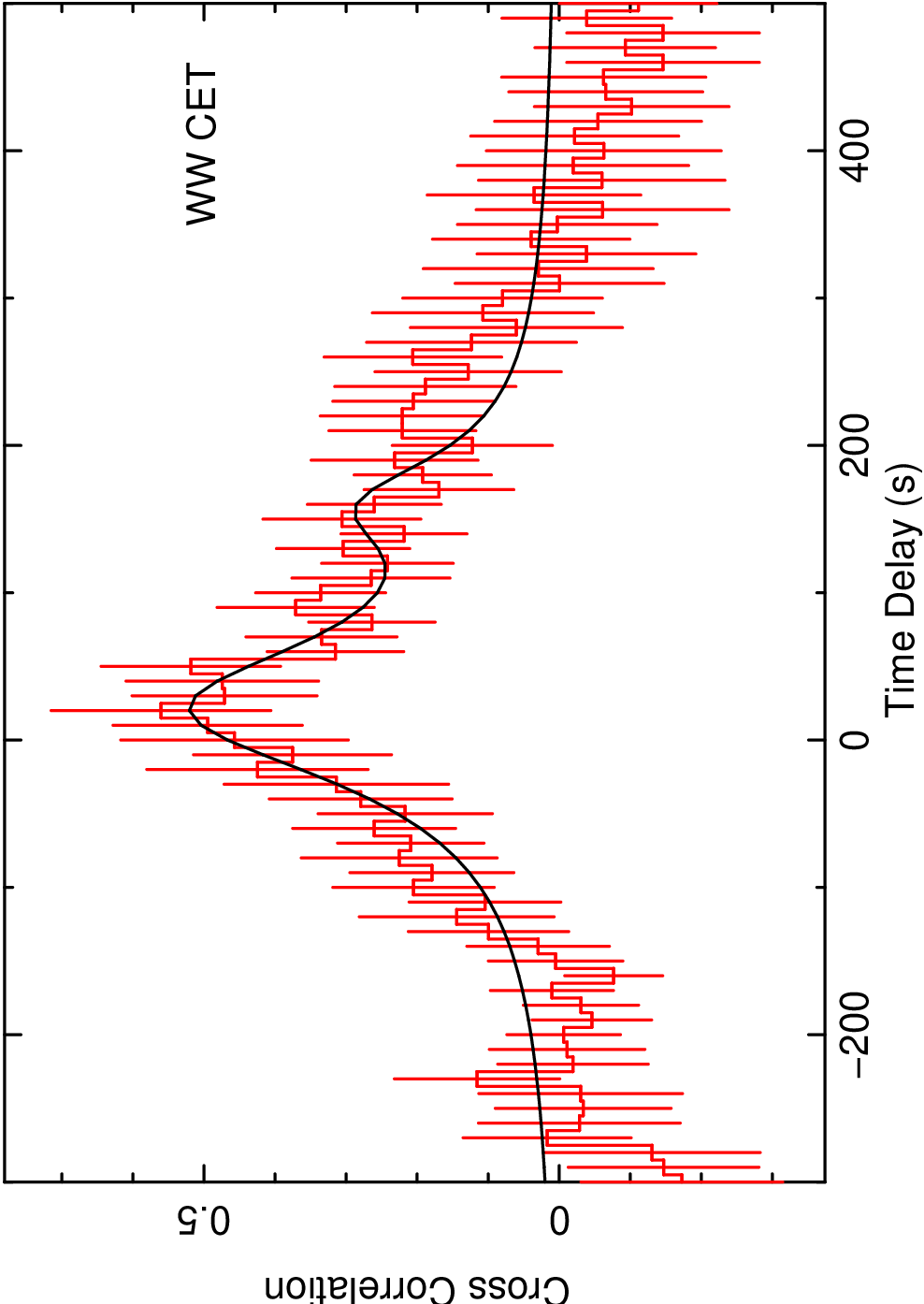}}
\resizebox{40mm}{!}{\includegraphics{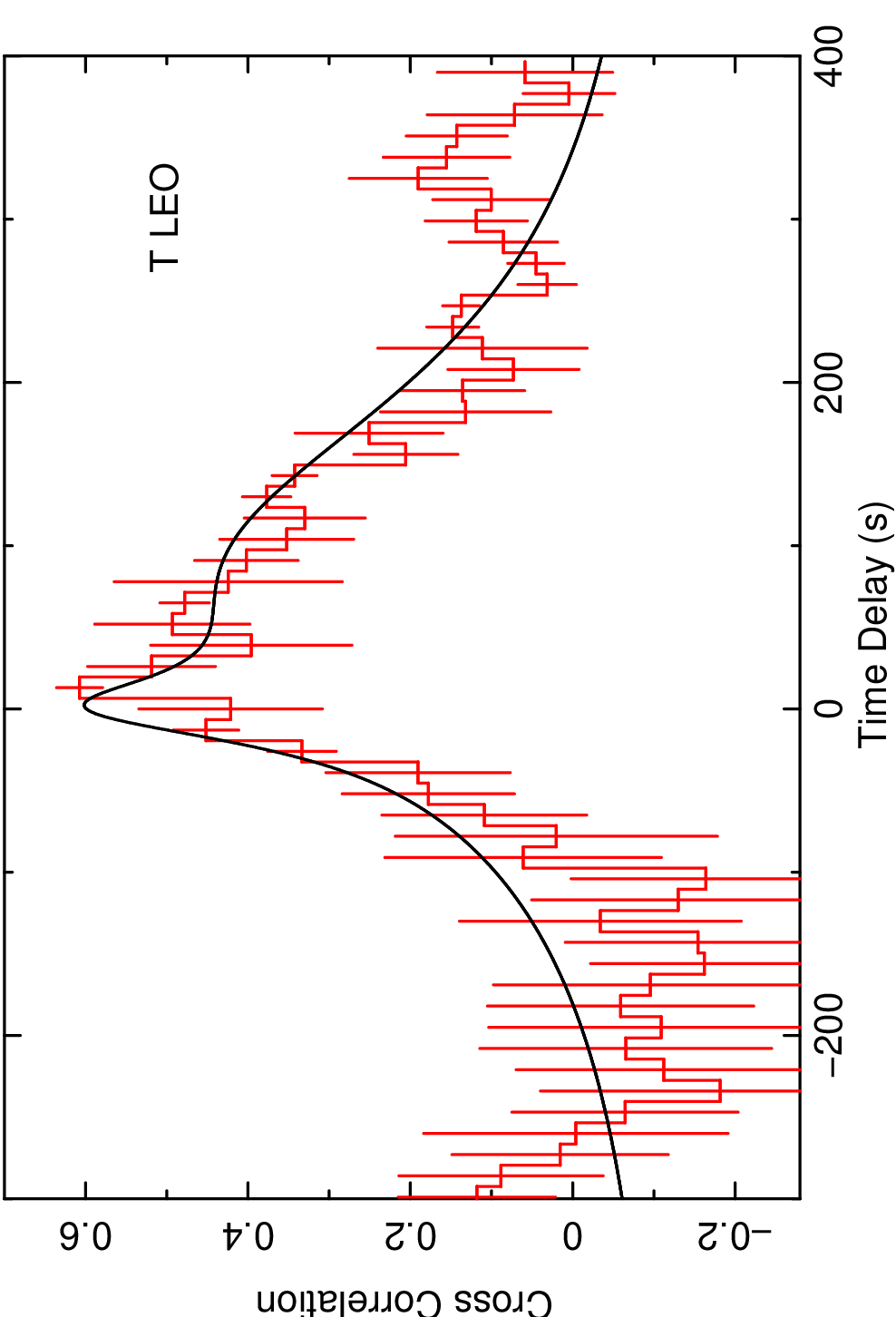}}
\resizebox{42mm}{!}{\includegraphics{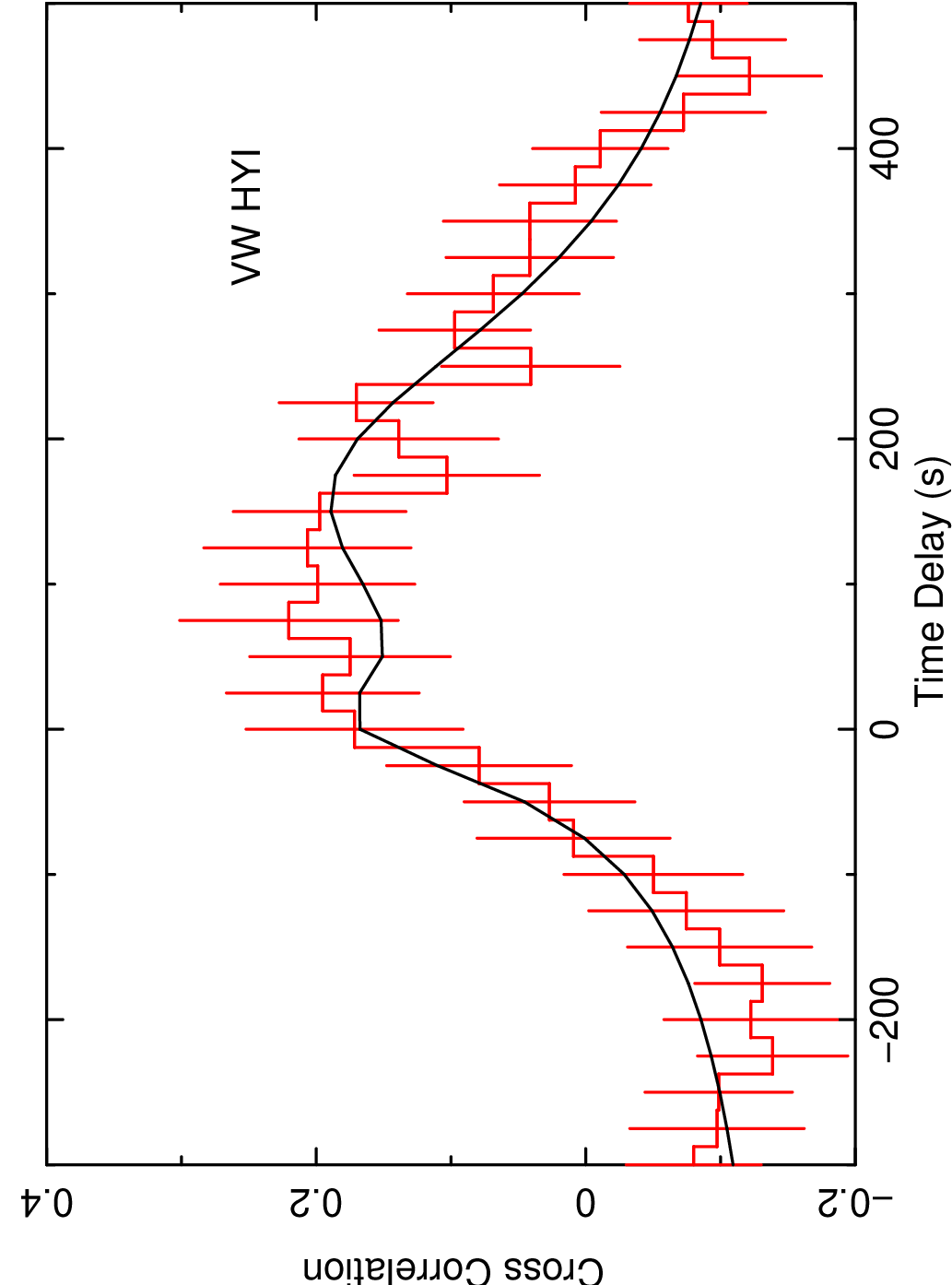}}

\end{turn}
\caption{ See Figure 10 in Balman \& Revnivtsev (2012).
The cross-correlation of the EPIC pn (X-ray) and OM (UV) light curves with 1 sec time resolution.
The two-component Lorenzian fits are shown as solid black lines
(except for SS Cyg).}
\end{myfigure}

\begin{myfigure}
\begin{turn}{-90}
\resizebox{50mm}{!}{\includegraphics{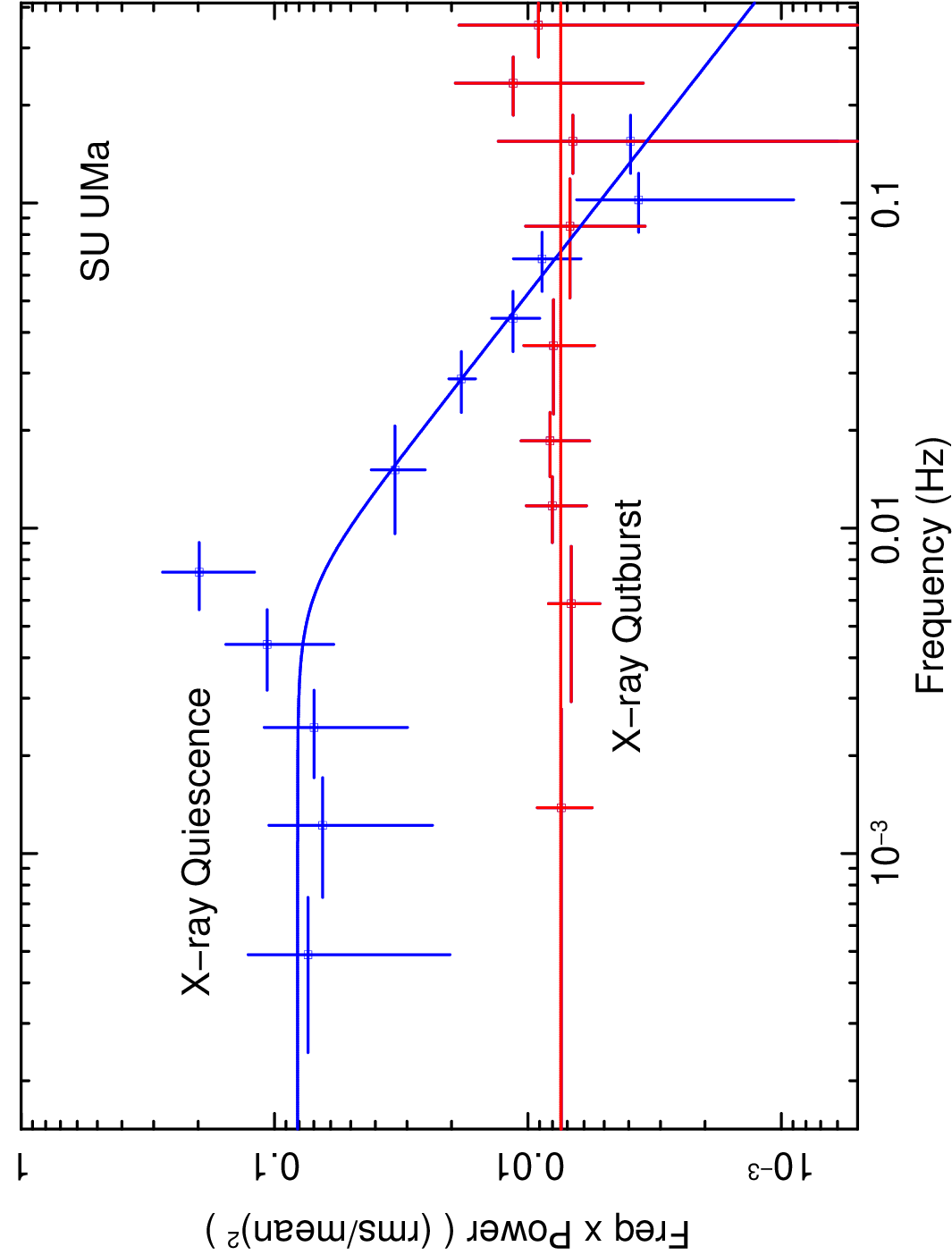}}
\resizebox{46mm}{!}{\includegraphics{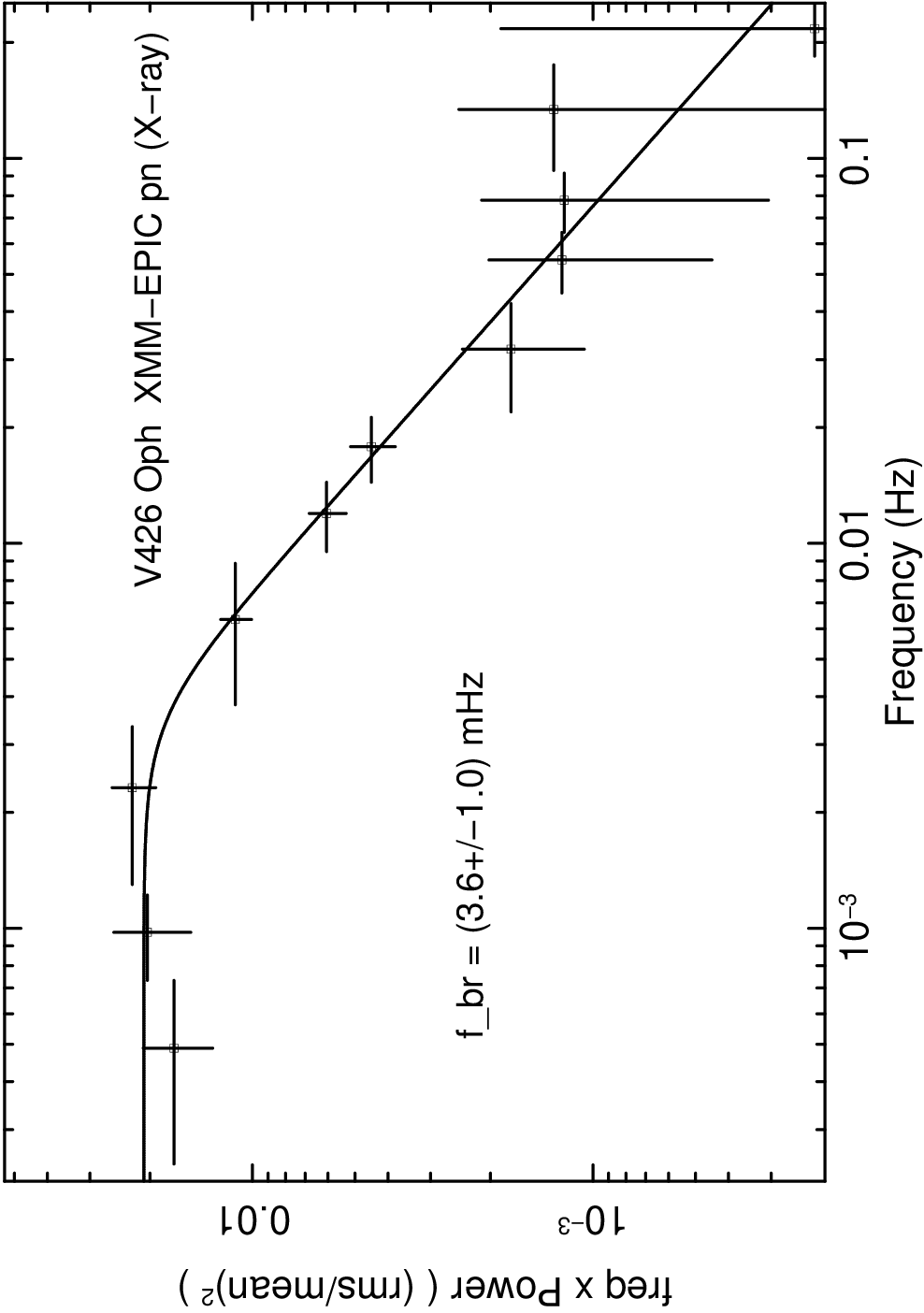}}
\resizebox{48mm}{!}{\includegraphics{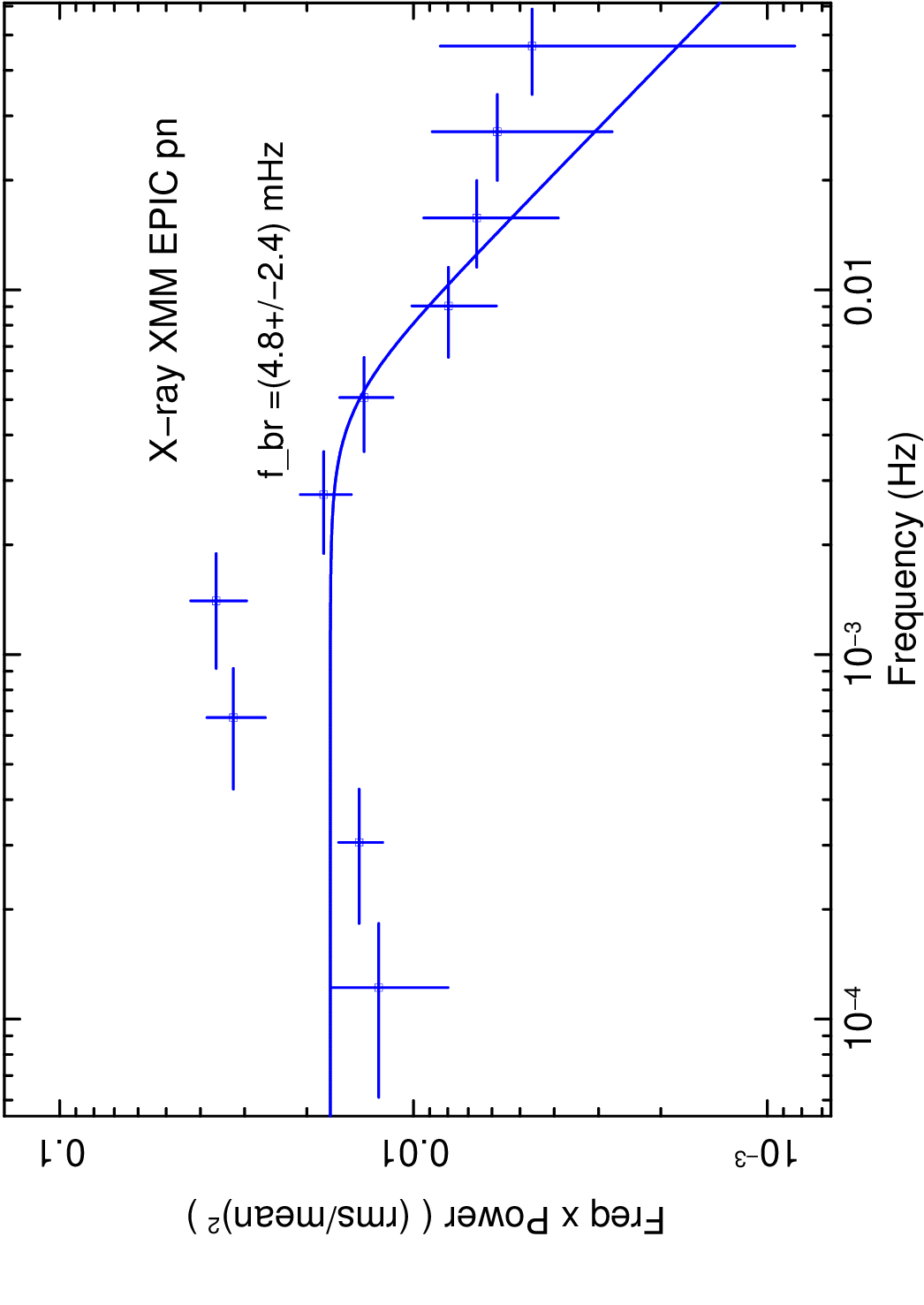}}
\end{turn}
\caption{
The $RXTE$ PDS of SU UMa in quiescence and in optical outburst is displayed (top panel).
$XMM$-$Newton$ EPIC pn PDS of V426 Oph and HT Cas in quiescence (middle and bottom panel).
The solid lines are the fits with the propagating fluctuations model. }
\end{myfigure}

An analysis on the $RXTE$ data of SU UMa in quiescence and outburst
following six consecutive outburts reveal a similar broad-band noise structure
to SS Cyg in quiescence showing a break frequency $\sim$ 5.5-7.5 mHz
with a truncated optically thick disk $\sim$ 3.8$\times$10$^{9}$ cm. The preliminary analysis
of the outburst data during the X-ray suppression episodes
(optical peak of the outburst) indicates no disk truncation or a truncation
around 0.1 Hz (see Figure 4).  Therefore, SU UMa indicates a similar behaviour
of the disk during the outburst to SS Cyg where the inner disk
moves in almost all the way to the WD during the optical peak
and moves out in decline to quiescent location further out.
The lower level of broad-band noise during outburst stage of SU UMa
may be due to the high radiation pressure support of an optically thick disk flow
during the peak of the outburst suppressing the variations.
A similar PDS analysis of the $XMM$-$Newton$ data of the DNe, V426 Oph and HT Cas in
quiescence reveals  break frequencies 4.6-2.6 mHz and 7.2-2.8 mHz, respectively.
The approximate Keplerian radii where the  optically thick disk truncates is
$\sim$ 6.2$\times$10$^{9}$ cm and $\sim$ 4.5$\times$10$^{9}$ cm, respectively (see Figure 4).

\section{Conclusions}

Studies of DNe broad-band noise characteristics in the X-rays (see also Balman \& Revnivtsev 2012) 
indicate that DNe have truncated accretion disks in quiescence 
detected in at least  8 systems in a range $\sim$(3.0-10.0)$\times$10$^9$\ cm including errors.
The Magnetic CVs (MCVs) show rather smaller truncation radii (0.9-2.0)$\times$10$^9$\ cm
(Revnivtsev et al. 2010, 2011). This
can also explain the UV and X-ray delays in the outburst stage and the accretion
may occur through hot (coronal)  
flows in the disk.
Note that extended emission and winds are detected from DN in the outburst stage
which may be an indication of the coronae/hot flows in these systems (e.g., Mauche 2004).
Time delays detected in a range of  96-181 sec, are also  
consistent with matter propogation timescales onto the WD in a truncated nonmagnetic CV disk
in quiescence.
Balman \& Revnivtsev (2012) approximate an $\alpha$ of 0.1-0.3 for the inner regions
of the DNe accretion disks in quiescence.

In addition, the outburst spectra of WZ Sge shows two component spectrum
of only hard X-ray emission, one of which may be fitted with a power law suggesting 
thermal Comptonization of the optically thick disk photons or scattering from
an existing wind during the outburst. The spectral evolution and disapearance 
of the power law component after outburst may support this issue and that the
accretion disk goes back to its quiescent truncated structure and Comptonization stops.

A general picture of the accretion flow around a WD in quiescence thus might be somewhat similar to
that of the black hole/neutron star accretors with an optically thick colder outer accretion disk
and an optically thin hot flow in the inner regions where the truncation occurs 
(see Esin et al. 1997, Done et al.
2007). The appearance of a hot flow (e.g., ADAF-like) in the innermost 
regions of the accretion disk will differ from
that of ordinary rotating Keplerian disk because it is no longer fully supported by rotation,
but might have a significant radial velocity component with sub-Keplerian speeds.  

It is important to monitor DNe in the X-rays to measure the
variability in the light curves in time together with the variations of the possible disk
truncation and
formation of plausible coronal (optically thin) and/or ADAF-like flows/regions on the disk in 
quiescence and outburst.

\thanks
SB thanks to M. Revnivtsev for his valuable collaboration on the timing analysis of DN.

\end{multicols}
\end{document}